\documentstyle[11pt,aaspp4,paper,psfig]{article}

\begin{document}

\title {\bf The RGB Sample of Intermediate BL\,Lacs}

\author{S.\ A.\ Laurent-Muehleisen \\
(slauren@igpp.llnl.gov)}
\affil{University of California-Davis and\\
The Institute for Geophysics and Planetary Physics,\\
Lawrence Livermore National Laboratory\\
7000 East Ave., Livermore, CA 94550, USA}
\authoraddr{L-413\\LLNL/IGPP\\7000 East Ave.\\Livermore, CA 94550}

\author{R.\ I.\ Kollgaard \\
(rik@fnal.gov)}
\affil{Fermi National Accelerator Laboratory, Batavia, IL 60510, USA}

\author{E.\ D.\ Feigelson \\
(edf@astro.psu.edu)}
\affil{Dept.\ of Astro.\ \& Astrophys., The Pennsylvania State University,\\
University Park, PA 16802, USA}

\author{W.\ Brinkmann and J.\ Siebert\\
(wpb@rzg.mpg.de, jos@mpe.mpg.de)}
\affil{Max-Planck-Institut f\"ur extraterrestrische Physik,\\
Giessenbachstrasse, D-85740, Garching, Germany}

\begin{abstract}

Combining newly identified and previously known BL\,Lacs from the RASS-Green
Bank (RGB) catalog, we present a sample of 127 BL\,Lacs, the largest ever
derived from a single uniform survey.  A Complete sample of 33 objects
brighter than O$=$18.0\,mag is also presented.  These samples are compared to
other known BL\,Lac samples and are generally found to exhibit properties
intermediate between those of the previously disparate classes of High and
Low energy peaked BL\,Lacs (HBLs and LBLs, respectively).  This result is most
dramatic in the distribution of the X-ray to radio logarithmic flux ratios,
where the RGB BL\,Lacs are shown to peak precisely where the sharp dichotomy
between the two subclasses was previously seen.  The $\alpha_{\rm ro}$ vs.\
$\alpha_{\rm ox}$ diagram also shows the RGB sample smoothly bridges the gap
between the previously distinct subclasses of LBLs and HBLs.  The range of
broadband Spectral Energy Distributions (SEDs) exhibited by the RGB objects
also shows that contrary to prior claims, searches based on relatively deep
surveys cannot limit followup spectroscopy to targets with a narrow range of
SEDs since BL\,Lacs clearly constitute a homogeneous population with a wide
range of SEDs.

Similar to results based on the EMSS and 1\,Jy BL\,Lac samples, we find a
weak, but statistically significant correlation between the composite X-ray
spectral index $\alpha_{\rm xox}$ and the radio-optical spectral index
$\alpha_{\rm ro}$.  This implies that the more LBL-like RGB BL\,Lacs have a
secondary source of X-ray emission, possibly from an inverse Compton
component.  This result, in addition to other characteristics of the RGB
sample, indicates that the simple unified scheme which postulates HBLs and
LBLs differ solely by orientation may be in need of revision.  We also present
both the X-ray and radio logN$-$logS distributions for which the competing
HBL/LBL unification scenarios have differing predictions.  The unknown effects
of the triple flux limit inherent in the RGB Complete sample makes
quantitative analysis uncertain, but the characteristics of the RGB sample
compare well both with results obtained from previous samples and with general
theoretical predictions based on a simple Monte Carlo simulation.  Our
analysis indicates that the unimodal distribution of BL\,Lac properties found
in the RGB sample likely reliably reflect the underlying population, while the
bimodal distribution found in earlier studies arose primarily from 
observational selection effects.  The presence of not only intermediate, but
also extreme HBL and LBL objects is the RGB survey's unique strength and
offers clear avenues for future studies which can undoubtedly address the
question of how HBLs and LBLs are related.

\end{abstract}


\clearpage

\section {Introduction}

BL\,Lacs comprise a rare subclass of Active Galactic Nuclei (AGN) and are
characterized by a lack of prominent emission lines, a highly variable
nonthermal continuum and strong, variable optical polarization
\cite[]{vista,unified}.  Additional characteristics include the lack of a
UV-excess (or ``blue bump'') and a core-dominated radio morphology
\cite[]{AS,WMA,lmheao}.  The dominance of nonthermal radiation at all
wavelengths ranging from the radio to gamma ray regimes is well-established
\cite[][and references therein]{rita} and makes BL\,Lacs particularly
interesting laboratories for the study of AGN phenomena.

Numerous studies have shown that BL\,Lacs contain relativistic jets which
produce narrow cones of beamed emission which makes the observed radiation
sensitive to orientation of the jet axis relative to the line-of-sight
\cite[see][and references therein]{vista,unified}.  BL\,Lacs are associated
with those objects that are oriented such that their jets lie close to the
line-of-sight while the parent population of misaligned objects are postulated
to be low luminosity Fanaroff-Riley Type\,I radio galaxies \cite[]{FR,browne}.
This association between BL\,Lacs and FR I galaxies is one of the stronger
links in the ``unified scheme'' of AGN where observed properties are primarily
a result of orientation rather than intrinsic astrophysical differences
\cite[e.g.,][]{USreview}. 

However, it has not been clear whether all BL\,Lac properties can be
attributed to orientation differences, nor whether BL\,Lacs constitute a
homogeneous class.  For example, the broadband spectral energy distributions
(SEDs) of BL\,Lacs discovered in X-ray and radio surveys differ significantly
which has led to the subclassification of BL\,Lacs into X-ray$-$selected and
radio$-$selected objects (XBLs and RBLs, respectively).  This has recently
been supplanted by a new classification ``High energy peaked BL\,Lacs'' (HBLs)
and ``Low energy peaked BL\,Lacs'' (LBLs) based on the  ratio of X-ray to
radio flux densities, S$_{\rm x}$/S$_{\rm r}$ \cite[]{GP,padovanigiommi}.
Generally, XBLs tend to be HBLs and exhibit less extreme properties than RBLs
which are usually LBLs.

The two BL\,Lac subclasses exhibit systematically distinct properties,
including the degree of radio core dominance, optical polarization fraction
and duty cycle, fraction of optical host galaxy light, and perhaps even
parsec-scale jet speeds, megaparsec-scale clustering properties and host
galaxy optical and radio luminosities
\cite[]{lmheao,perlman,jannuzi,hostgal,ronvlbi1,ronvlbi2}.  Many of these
characteristics are consistent with the unified scheme paradigm if HBLs are
objects which lie further from the line-of-sight than LBLs.  However, 
orientation by itself cannot explain the apparent lack of BL\,Lacs with
properties intermediate between the LBL and HBL subclasses.

There is also increasing concern that the simplest unified scheme may not
account for all subclass distinctions.  Intrinsic as well as orientation
differences may be present.  For example, estimates of jet speeds and angles
to the line-of-sight of HBLs, LBLs and FR\,I radio galaxies do not appear to
be able to account for all differences in the SEDs \cite[]{rita}.  In
addition, estimates of space densities are inconsistent with orientation
values, and the cosmic evolution of the classes appears to be incompatible
\cite[]{travis}.

These issues have proved difficult to address because existing BL\,Lac samples
are still relatively small and were generated from shallow surveys which 
contain only the very brightest objects in either the radio or X-ray
wavebands.  These and other selection effects have produced samples biased
towards the most extreme HBLs or LBLs with few transitional objects.  The
sample of BL\,Lacs presented here was generated from a cross-correlation of a
deep radio \cite[]{rgb} and X-ray catalog and contains BL\,Lacs with the full
range of properties from HBLs to LBLs.  This RASS-Green Bank (RGB) BL\,Lac
sample is the largest BL\,Lac sample yet created from a uniformly defined set
of criteria.  It consists of 127 objects drawn from a correlation of the ROSAT
All-Sky Survey (RASS) and a reanalysis of the 1987 Green Bank 6\,cm radio
survey \cite[GB96,][]{gnc2}.  The design and followup spectroscopic
observations of this sample are presented in \nocite{rgbIDs}
Laurent-Muehleisen et al.\ (1998, hereafter Paper~I).  Here we concentrate
specifically on the RGB BL\,Lacs and on what they reveal about the
relationship between the BL\,Lac subclasses.

This paper is organized as follows.  In \S\ref{candidate_selection} we briefly
review the RASS/GB correlation and our followup VLA and optical observations.
The RGB and ``RGB Complete'' samples are presented in \S\ref{rgb_sample} and
\S\ref{rgb_complete}.  Sections \ref{properties} and \ref{unified_scheme}
analyze these samples' bulk characteristics and discuss their astrophysical
implications.  We assume throughout H$_{\rm o}$=100 km\,s$^{-1}$\,Mpc$^{-1}$,
q$_{\rm o}$$=$0.5 and define spectral indices, $\alpha$, such that ${\rm
S}_{\nu} \propto \nu^{-\alpha}$.

\section{Selection of Candidate Objects}\label{candidate_selection}
Both radio and X-ray surveys have proven to be a rich source of new BL\,Lacs.
The largest purely radio$-$selected sample is that based on the 1\,Jy survey
while the most prominent X-ray$-$selected samples are those based on the
Einstein Extended Medium Sensitivity Survey
\cite[EMSS;][]{gioia,stocke,maccacaro} and the HEAO-1 Large Area Sky Survey
\cite[]{wood,heao}.  As the flux limit and/or sky coverage of radio and X-ray
surveys has improved and the ability to fully identify all objects in these
surveys has become impractical, the technique of selecting candidate BL\,Lacs
based on the broadband SEDs of previously known BL\,Lacs has proven to be
highly efficient.  The Einstein Slew Survey \cite[]{elvis,SES,slew}, Hamburg
Quasar Survey \cite[]{hqs,NA}, Deep X-ray Radio Blazar Survey samples
\cite[]{dxrbs} and the optical polarization sample of \cite{KO} were all
created by spectroscopically classifying sources with counterparts detected
concurrently in the more than one band.  Nevertheless, the number of objects
in each of these samples has remained relatively small and their selection
effects are often difficult to assess \cite[see, e.g.,][]{marcha1}.  The need
for a large sample with a minimal number of simple selection criteria is
clear.

The potential of the ROSAT All-Sky Survey for creating just such a BL\,Lac
sample has been noted \cite[]{stocke89} and followup programs are confirming
this prediction.  Some programs based on the RASS have used the optical
polarization or narrowly defined broadband SEDs to select BL\,Lac candidates
\cite[]{KO,NA}.  These candidates were chosen independent of any knowledge of
the radio flux of the source, but followup observations have shown that all
are detected at centimeter wavelengths \cite[]{NA}, a result in agreement with
the assertion that radio-silent BL\,Lacs are very rare or possibly nonexistent
\cite[]{radioquiet}.  Our RGB BL\,Lac sample was therefore constructed using
previously detected radio emission as a criterion.  This sample is therefore
triply flux-limited (radio, optical and X-ray) but imposes no other selection
criteria other than location in the northern hemisphere
\cite[0$^{\circ}$$<$$\delta$$<$$75^{\circ}$;][]{rgb}.  This method coupled
with high sensitivity of the RGB survey in these three wavebands detects
BL\,Lacs with a variety of broadband SEDs.

The initial RGB catalog consisted of sources whose positions differed by less
than 100$\arcsec$ in the RASS and a point source catalog created from the 1987
Green Bank 6\,cm radio survey \cite[GB96,][]{gnc2,neumann}.  This new GB
catalog consists of 3$\sigma$ and greater confidence sources and has a flux
density limit of $\sim$15\,mJy in the declination range from
$30^{\circ}-75^{\circ}$ and increases to $\sim$24\,mJy at low declinations
\cite[]{neumann}.  In order to eliminate spurious RASS-GB coincidences and to
determine positions accurate enough for reliable optical identification, the
2,127 sources in the RASS-GB correlation were observed at high resolution with
the NRAO's\footnote{NRAO is operated by Associated Universities, Inc., under
cooperative agreement with the National Science Foundation.} VLA.  Two radio
catalogs were produced, the first consists of 1,861 sources for which
subarcsecond positions and core radio flux densities were obtained; the second
consists of 436 sources for which only low resolution data ($\sim$8$\arcsec$
positional accuracy) were obtained \cite[]{rgb}.  Our analysis showed that all
sources whose radio and X-ray positions differed by less than 40$\arcsec$ are
true matches to a high degree of confidence.  Additional details on the
RASS-GB correlation and the followup radio observations can be found in
\cite{rgb} and \cite{thesis}\footnote{This Ph.D.\ thesis is available via the
WWW at http://www-igpp.\linebreak[0]llnl.\linebreak[0]gov/\linebreak[0]people/
\linebreak[0]slauren.html}.

Optical counterparts were determined via Automatic Plate Measuring scans of
the high Galactic latitude ($>$25$^{\circ}$) POSS\,I photographic plates
\cite[]{apm}.  Optical counterparts within 3$\arcsec$ of RGB sources were
identified and both the O (blue) and E (red) magnitudes measured
\cite[]{thesis,siebert}.  A looser criterion of 5$\arcsec$ was used for
sources in the low resolution VLA catalog.  Spectra were obtained for 169
optically bright (O$\le$18.5\,mag) objects which lacked spectroscopic
classifications (Paper I).

Table \ref{tab:survey_numbers} summarizes the steps involved in the creation
of the final RGB catalog.  A discussion of the broadband multiwavelength
properties of the entire RGB sample can be found in Paper~I and \cite{siebert}
as well as \cite{thesis}.

\placetable{tab:survey_numbers}
\begin{table}
\dummytable \label{tab:survey_numbers}
\end{table}
\begin{figure}[t]
\vspace*{-4.10in}
\hspace*{-1.00in}
\psfig{file=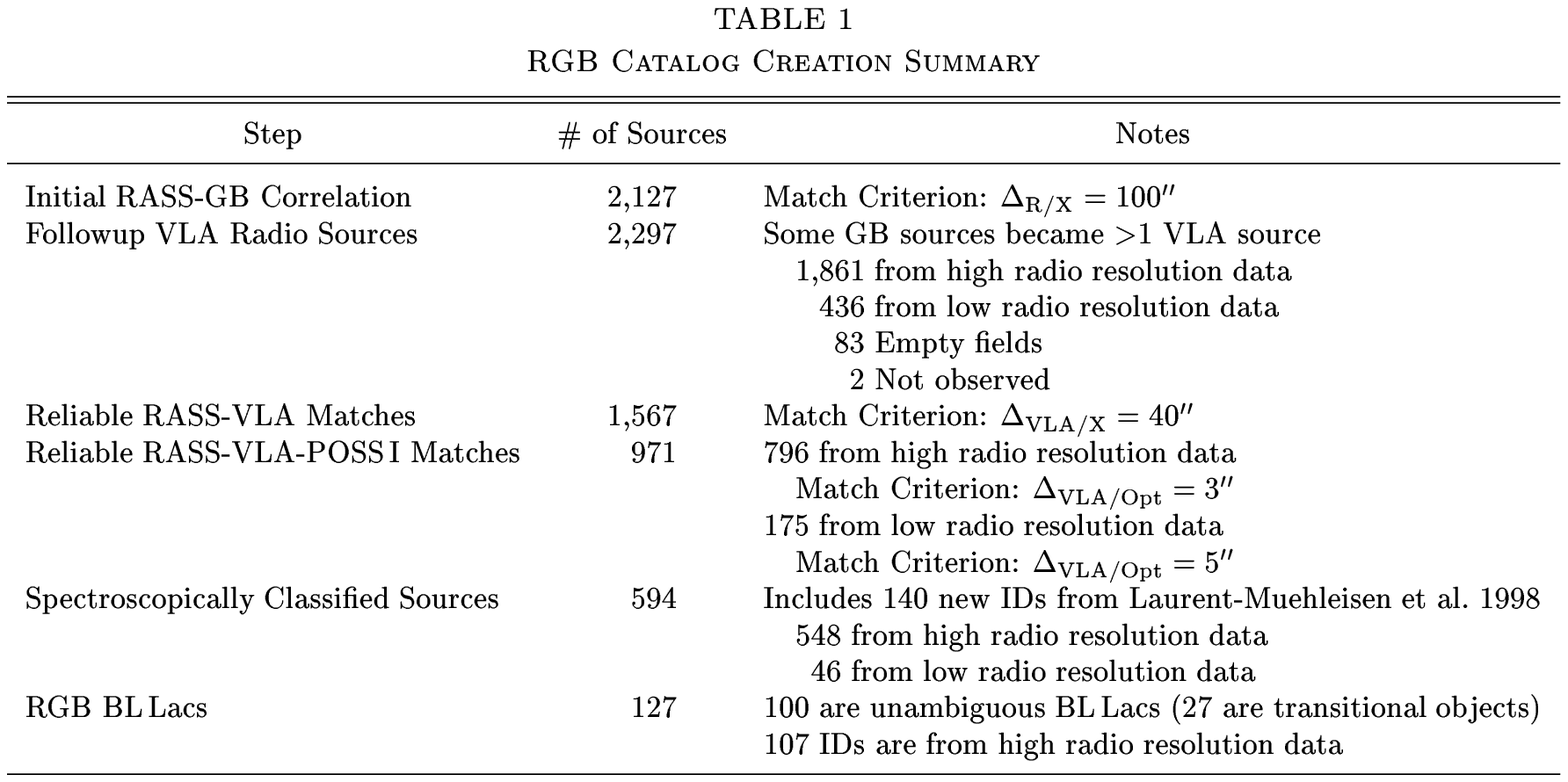,height=11.0in,width=8.5in}
\end{figure}
\clearpage

We noted in Paper~I that the operational definition of a BL\,Lac has changed
since that of \cite{stein1} who defined a BL\,Lac as an AGN having a highly
variable, linearly polarized, nonthermal continuum without optical emission
lines.  More recently, \cite{stocke} defined a BL\,Lac as an AGN with emission
lines whose equivalent width does not exceed 5\,\AA\ and whose
\ion{Ca}{2}~H\&K break strength (Br$_{4000}$) is $\le$25\%\footnote{The
\ion{Ca}{2} break contrast refers to the relative depression of the continuum
blueward of the \ion{Ca}{2}~H\&K lines (3933\,\AA\ \& 3968\,\AA).}.  However,
recent observations, particularly of the less extreme HBLs, has shown that
even this definition is too stringent and excludes objects which otherwise
exhibit BL\,Lac-like properties and should therefore be classified as such
\cite[]{marcha2,scarpa}.  Establishing a set of criteria which define the
BL\,Lac class is particularly difficult because much of the observed emission
is contributed by orientation-dependent beamed radiation while the creation of
unbiased samples requires that the classification of objects be based on
intrinsic (not observed) characteristics.  In Paper~I, we adopted the
classification criteria proposed by \cite{marcha2} to distinguish BL\,Lacs
from ordinary quasars, Seyfert, radio and elliptical galaxies.  Briefly, our
classification scheme is:

\begin{itemize}
\item{If the spectrum is featureless or the only features observed are
emission lines with W$_{\lambda}$$\le$5\,\AA\ (rest frame), the object is
classified as a BL\,Lac.}

\item{If absorption features are present and Br$_{4000}$$<$25\%, we classify
the object as a BL\,Lac, provided any emission lines present have
W$_{\lambda}$$\le$5\,\AA.}

\item{If the \ion{Ca}{2} break contrast is between 25\%$-$40\%, we classify
the object as a possible BL\,Lac if any emission line present also has an
equivalent width smaller than that required by the \cite{marcha2} criterion
for that particular break strength (see Paper~I).}

\item{If the \ion{Ca}{2} break contrast was $>$40\% we classified the object
as a galaxy because of the lack of spectroscopic evidence for an AGN, although
we find no clear discontinuity in properties which distinguishes BL\,Lacs from
galaxies (Paper~I, Figure 3).}
\end{itemize}

Our spectroscopic observations combined with these criteria produced a sample
of 53 RGB BL\,Lacs, 38 of which were newly discovered.  We now combine these
objects with previously identified BL\,Lacs in the RGB catalog and discuss the
full sample.

\section{The RGB Sample of BL\,Lacs}\label{rgb_sample}
The RGB sample consists of 127 sources of which 100 are definitive BL\,Lacs.
Many of the 27 objects which are only probable BL\,Lacs have break contrasts
larger than 25\%, but adhere to the March\~{a} Br$_{4000}$$-$W$_{\lambda}$
criteria.  Twenty of the RGB BL\,Lacs belong to the low radio resolution
subset of the RGB catalog.  Because core radio flux characterizes many of the
beaming properties of BL\,Lacs, we exclude these ``low radio resolution
objects'' from all further analysis, reducing the sample of objects we discuss
to 107 RGB BL\,Lacs which is the largest sample of BL\,Lacs ever cataloged
from one survey.

For completeness, all 127 sources are presented in Table \ref{tab:rgb_bllac}.
The columns give the (1) RGB Name; (2) alternate Name; (3) and (4) J2000
coordinates; (5) RGB 5\,GHz core radio flux density (in mJy); (6) X-ray flux
(in $10^{-12}$ erg\,s$^{-1}$cm$^{-2}$ in the 0.1-2.4\,keV ROSAT band); (7) O
magnitude from the APM POSS-I catalog; (8) ROSAT PSPC X-ray spectral energy
index; (9) and (10) the $\alpha_{\rm ro}$ and $\alpha_{\rm ox}$ spectral
indices; (11) redshift and (12) references.  As discussed in \cite{siebert},
\cite{rgb} and \cite{rgbIDs}, the uncertainties on the multiband fluxes can
be taken to be approximately 20\%, 0.5\,mag and 25\% in the radio, optical
and X-ray, respectively.  \nocite{lamer} \nocite{perlmanxbls} \nocite{KO}
\nocite{rosatrbls} \nocite{GHI} \nocite{KWRG} \nocite{fleming}
\nocite{badeBLL} \nocite{B98} \nocite{Brinkmann} \nocite{rgb} \nocite{rgbIDs}
\nocite{NA} \nocite{PGcat} \nocite{siebert} \nocite{VTRB} \nocite{marcha2}
\nocite{henstock} \nocite{brightrosatIDs} \nocite{beckmann}

Flux densities in Table \ref{tab:rgb_bllac} have not been K-corrected, but the
two-point spectral indices ($\alpha_{\rm ro}$ and $\alpha_{\rm ox}$) are
corrected assuming a flat ($\alpha_{\rm r}$$=$0.0) radio spectral index and an
X-ray spectral index as given in column 8, or the mean X-ray spectral index of
$\alpha_{\rm x}$$=$1.2, valid for a large sample of RASS-detected BL\,Lacs
\cite[]{siebert}.  When a measured redshift is unavailable, the K-correction
is based on the median RGB BL\,Lac redshift, z$=$0.16.  Although this probably
underestimates these objects' true redshift (objects with z$\leq$0.16 would
likely show spectral evidence for the host galaxy and hence would have a
measured redshift), we feel it is important to apply some type of K-correction
to all objects since the K-correction always produces a net flattening of the
$\alpha_{\rm ro}$ and $\alpha_{\rm ox}$ spectral indices (given our
assumptions).  However, the effect is small, typically producing an error of
$\le$0.05 in either index unless the true redshift is $>$1.0.

The optical magnitudes are equivalent O band values, converted from other
bands assuming $\alpha_{\rm opt}$$=$1.0 when no O or B magnitudes were
available in the literature.  (The notes to Table \ref{tab:rgb_bllac} also
give the original magnitude and band.)  If any galaxy absorption features were
detected, we corrected the magnitudes to include only the contribution from
the AGN component using the method described in Paper~I.  We note that these
corrections have only been made for the newly identified objects presented in
Paper~I because the correction requires the determination of the \ion{Ca}{2}
break contrast which is usually not reported in the literature.

The K-corrected radio$-$optical and optical$-$X-ray spectral indices are
defined to be $\alpha_{\rm ro}$$=$$0.194\,\log \left (\frac{{\rm S}_{5}^{\rm
core}}{{\rm S}_{\rm opt}} \right )$ and $\alpha_{\rm ox}$$=$$0.351\,\log
\left(\frac{{\rm S}_{\rm opt}}{{\rm S}_{\rm X}} \right )$ and are given in
columns 9 \& 10.  The monochromatic ROSAT X-ray flux density was converted
from the fluxes given in Table \ref{tab:rgb_bllac} according to:\\

\begin{equation}
{\rm S}_{\rm X} = 4.138\times10^{-18} \cdot {\rm F}_{\rm X} \cdot {\rm
E}^{-\alpha_{\rm x}} \left[ \frac{1-\alpha_{\rm x}}{2.4^{(1-\alpha_{\rm
x})}-0.1^{(1-\alpha_{\rm x})}} \right]
\end{equation}

\noindent where E is set at 2\,keV, F$_{\rm X}$ is in erg\,s$^{-1}$\,cm$^{-2}$
and S$_{\rm X}$ is in erg\,s$^{-1}$\,cm$^{-2}$\,keV$^{-1}$.  The values of
$\alpha_{\rm ro}$ and $\alpha_{\rm ox}$ are accurate to $\pm0.1$ and $\pm0.2$,
respectively, taking into account the observational uncertainties in the
fluxes.  This does not, however, include the effects of real temporal
variability in these nonsimultaneous multiband data.  Generally speaking,
variability is unlikely to change the flux values by more than 25\% and the
two-point spectral indices given in the table are therefore reasonably
representative.

Finally we note that a few RGB sources identified as BL\,Lacs in the
literature are missing from Table \ref{tab:rgb_bllac} and therefore
technically not members of the RGB BL\,Lac sample.  RGB\,J1058$+$564,
RGB\,J1110$+$715, and RGB\,J1610$+$671B have spectra that were presented in 
Paper~I, but violate the radio/X-ray coincidence criterion of $\Delta_{\rm
rx}$$\le$40$\arcsec$.  The previously known BL\,Lacs RGB\,J0738$+$177
(PKS\,0735$+$178) and RGB\,J1508$+$271 are respectively $3.1\arcsec$ and
$3.8\arcsec$ from bright optical sources and therefore do not appear in Table
\ref{tab:rgb_bllac} since they violate our radio/optical coincidence
criterion.  However, both are likely true RGB BL\,Lacs\footnote{Source
0930$+$4950 is a BL\,Lac but was incorrectly reported in \cite{rgb} as
belonging to the RGB catalog of X-ray and radio-emitting AGN.  We therefore
also exclude this object from the current paper.}.  \nocite{Brinkmann}
\nocite{NA} \nocite{KO} \nocite{PGcat} \nocite{brightrosatIDs}

\placetable{tab:rgb_bllac}
\begin{table}
\dummytable \label{tab:rgb_bllac}
\end{table}

\section{The RGB Complete Sample}\label{rgb_complete}
\begin{figure}[t]
\hspace*{1.0in}
\psfig{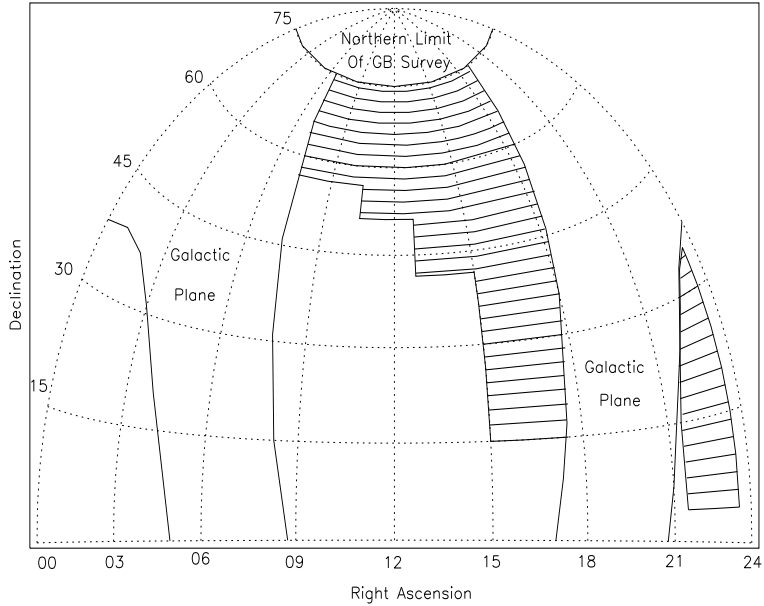}
\caption[]{The 3970\,deg$^2$ from which the RGB Complete sample was selected.
The Galactic plane (b$<$25$^{\circ}$) and northern extent of the GB survey
($\delta$$=$75$^{\circ}$) are labeled.  The region designated by the
horizontal lines constitutes the RGB Complete Survey area which consists of
the following six regions: (1) 6$^{\rm h}$30$^{\rm m}$$< \alpha <$10$^{\rm
h}$31$^{\rm m}$ and 57$^{\circ}$40$^{\prime}$$< \delta <$75$^{\circ}$ (2)
10$^{\rm h}$31$^{\rm m}$$< \alpha <$12$^{\rm h}$48$^{\rm m}$ and
50$^{\circ}$$< \delta <$75$^{\circ}$ (3) 12$^{\rm h}$48$^{\rm m}$$< \alpha
<$15$^{\rm h}$ and 40$^{\circ}$30$^{\prime}$$< \delta <$75$^{\circ}$ (4)
15$^{\rm h}$$< \alpha <$19$^{\rm h}$ and 15$^{\circ}$$< \delta <$75$^{\circ}$
and (5) 21$^{\rm h}$30$^{\rm m}$$< \alpha <$23$^{\rm h}$30$^{\rm m}$ and
3$^{\circ}$$< \delta <$34$^{\circ}$ with the added constraint that
b$>$25$^{\circ}$.\label{fig:coverage}}
\end{figure}

The RGB sample was constructed without imposing any selection criteria other
than the presence of an optical counterpart within 3$\arcsec$ (5$\arcsec$ for
the low radio resolution sample) and a RASS source within 40$\arcsec$ of a
GB96 radio source.  These well-defined criteria allowed us to create the ``RGB
Complete Sample'' which consists of optically bright objects (total O
magnitude $<$18.0) observed over 3970\,deg$^2$ of the sky with a completeness
of 94\% (Figure \ref{fig:coverage}).

\placefigure{fig:coverage}

Our complete survey area contains a total of 183 RGB sources, most of which
are emission-line AGN (68\%) and galaxies (12\%).  But 33 are optically bright
BL\,Lacs (including three candidate objects with \ion{Ca}{2} break contrasts
between 29-39\%) which constitute the RGB Complete BL\,Lac Sample.  These
objects are listed separately in Table \ref{tab:rgb_bllac}.  The RGB Complete
sample is therefore flux-limited in three bands: radio, optical and X-ray.
The constraints imposed by the RASS survey (whose flux limit varies with
ecliptic latitude and ${\em N}_{\rm H}$) and the GB survey (whose flux density
limit varies slightly with declination) both affect the catalog's overall
completeness but in a well-defined manner.  The effects of the optical flux
limit are more serious than those imposed by the X-ray and radio limits.
Assuming all BL\,Lacs have color-color indices in the range 0.1$<$$\alpha_{\rm
ro}$$<$0.8 and 0.5$<$$\alpha_{\rm ox}$$<$2.1 (see Figure \ref{fig:aro_aox}), a
limiting optical magnitude of O$=$18.0\,mag implies that the RGB Complete
sample is only truly complete above S$_{\rm r}$$=$3\,Jy and F$_{\rm
X}$$=$10$^{-11}$erg\,s$^{-1}$cm$^{-2}$, criteria which are satisfied by none
of the BL\,Lacs in this sample.  However, the situation is less dire in
practice since the RGB Complete sample is constructed from surveys with 
well-defined flux limits but without any additional selection criteria applied
to it.  Any additional constraints (such as limiting candidates to objects
with particular SEDs), however, would be much less useful for characterizing
the full BL\,Lac population.

We note that three previously known BL\,Lacs nearly, but do not exactly,
satisfy the criteria for inclusion in the RGB Complete sample.  First,
1ES\,2326$+$174 is 16.8\,mag Slew Survey BL\,Lac with 27\,mJy radio flux and
$1.5 \times 10^{-11}$ erg s$^{-1}$ cm$^{-2}$ X-ray flux \cite[]{slew}.  This
source lies in a region of the All-Sky Survey where the standard data
screening software finds very little acceptable data, reducing the effective
exposure time to essentially zero.  This source therefore did not appear in
our original RASS$-$GB correlation and is therefore missing from the current
RGB BL\,Lac sample.  Second, as mentioned in \S\ref{rgb_sample},
RGB\,J1058$+$564 and RGB\,1508$+$271 are BL\,Lacs that respectively violate
our radio/X-ray and radio/optical offset criteria.  With these possible
exceptions, our 33 source optically bright RGB BL\,Lac sample is complete
given our selection criteria and therefore constitutes a useful sample for
statistical study.

\section{Properties of the RGB Sample}\label{properties}
To assess the characteristics of the RGB sample, we compare it with four
large, well-defined samples of BL\,Lacs currently available in the literature:
the 1\,Jy, HEAO, EMSS and Einstein Slew Survey samples
\cite[]{stickel,stocke,slew,remillard,schwartz}.  We will treat the RBL 1\,Jy
sample as representing the range of properties associated with LBLs (Low
energy peaked BL\,Lacs), although the recent assertion that some 1\,Jy objects
may be misclassified microlensed sources should be noted \cite[]{mg2}.  The
XBL EMSS and HEAO samples will be combined to represent HBLs (High energy
peaked BL\,Lacs).  As will be shown below, the Slew Survey objects tend to
exhibit properties intermediate between HBLs and LBLs and are in that sense
similar to the RGB sample.   The occassional duplication of a source in more
than one sample is ignored here.

We compare the distributions of redshift, three representations of SED shape
$-$ S$_{\rm x}$/S$_{\rm r}$ ratio, $\alpha_{\rm xox}$$=$$\alpha_{\rm
ox}$$-$$\alpha_{\rm x}$ \cite[]{rita} and location in the $\alpha_{\rm ro}$
vs.\ $\alpha_{\rm ox}$ plane $-$ and the radio and X-ray BL\,Lac logN$-$LogS
relations.  Data for the comparison samples were obtained from the literature,
mainly from \cite{lmheao}, \cite{heao}, \cite{rita}, \cite{slew},
\cite{perlmanxbls}, \cite{rosatrbls}, \cite{siebert} and Paper~I.  The median
values for the various samples are given below in Table \ref{tab:median} and
were calculated using the Astronomy SURVival analysis software \cite[ASURV,
Rev.\ 1.2, ][which can be obtained from
http://www.astro.psu.edu/statcodes]{asurv} which properly handles the upper
and lower limits present in the data.  (One source has an upper limit to its
core radio flux density, RGB\,J1000$+$225A.)  Median values are calculated
using the maximum-likelihood Kaplan-Meier estimator and the statistical
significance of any differences between two samples is estimated with the
logrank and Peto \& Peto generalized Wilcoxon tests \cite[]{kmest}.

\placetable{tab:median}
\begin{deluxetable} {lllllllll}
\tablenum{3}
\tablecolumns{9}
\tablewidth{0pt}
\tablecaption{Median Properties of BL~Lac Samples\label{tab:median}}
\tablehead{
\colhead{} &
\colhead{HEAO} &
\colhead{EMSS} &
\colhead{XBL} &
\colhead{Slew} &
\colhead{RGB} &
\colhead{RGB} &
\colhead{1\,Jy}\\
\colhead{Property} &
\colhead{} &
\colhead{} &
\colhead{(HEAO$+$} &
\colhead{} &
\colhead{Complete} &
\colhead{}\\
\colhead{} &
\colhead{} &
\colhead{} &
\colhead{ EMSS)} &
\colhead{} &
\colhead{} &
\colhead{} &
\colhead{} }
\startdata
Redshift&\phm{$-$}0.12&\phm{$-$}0.30&\phm{$-$}0.20&\phm{$-$}0.16&\phm{$-$}0.13&\phm{$-$}0.16&\phm{$-$}0.50\nl
$\log$ S$_{\rm x}$/S$_{\rm r}$& $-$4.78 & $-$4.84 & $-$4.79 & $-$4.56 & $-$5.56 & $-$5.61 &     $-$6.89   \nl
$\alpha_{\rm xox}$            & $-$0.25 & $-$0.13 & $-$0.17 & $-$0.23 & $-$0.11 & $-$0.14 & \phm{$-$}0.23 \nl
\enddata
\end{deluxetable}

\begin{figure}[t]
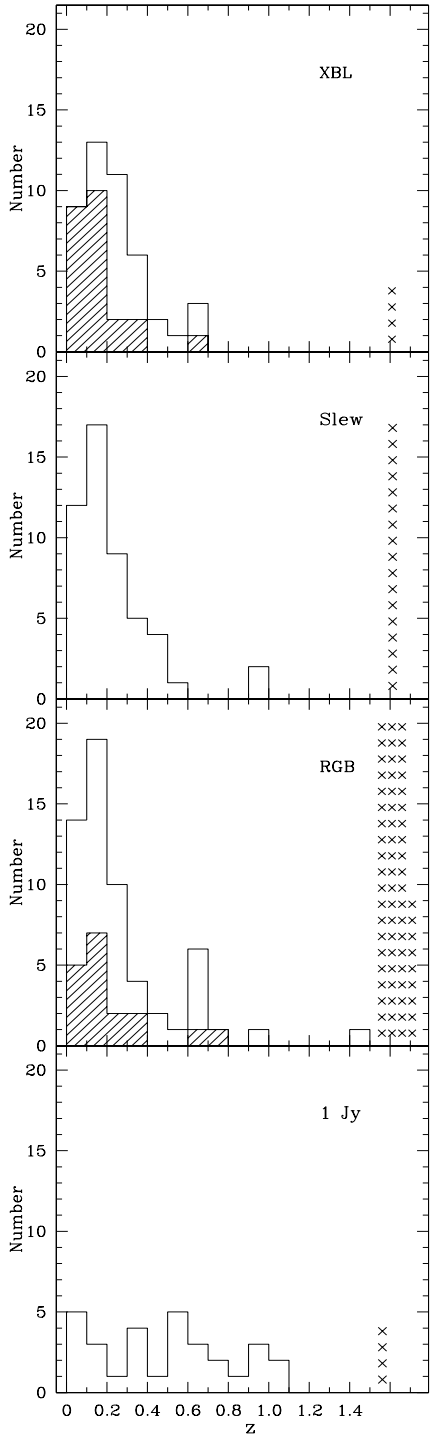

\vspace*{-1.3cm}
\psfig{file=xbl.z.epsi,height=5.5cm,width=5.9cm}
\vspace*{-0.915cm}
\psfig{file=slew.z.epsi,height=5.5cm,width=5.9cm}
\vspace*{-0.92cm}
\psfig{file=rgb.z.epsi,height=5.5cm,width=5.9cm}
\vspace*{-0.92cm}
\psfig{file=1jy.z.epsi,height=5.5cm,width=5.9cm}
\vspace*{-0.46cm}
\caption[]{The redshift distribution for the various BL\,Lac samples. {\bf a.}
The XBL sample of BL\,Lacs which consists of both the EMSS and HEAO objects.
The hatched region denotes the HEAO objects and the top solid line is the sum
of both samples. {\bf b.} The Slew survey sample of BL\,Lacs. {\bf c.} The RGB
sample of BL\,Lacs. The hatched region denotes the RGB Complete sample of
bright objects.  {\bf d.} The 1\,Jy sample of RBLs.  The $\times$'s on the far
right represent objects with unknown redshifts.\label{zfig}}
\end{figure}

\placefigure{zfig}
\clearpage

\subsection{Redshift Distribution}\label{subsect:z}
Figure \ref{zfig} shows the distribution of redshifts for the RGB sample as
well as the XBL, Slew Survey, and 1\,Jy samples.  A typical HBL clearly
resides at a lower redshift than a typical LBL, a trend which is predicted by
the unified model of \cite{bolometricL} which asserts that HBLs constitute the
intrinsically lower luminosity sources.  Although redshifts are known for only
approximately half (59) of the RGB sources, RGB redshifts span nearly the
entire range exhibited both by HBLs and LBLs but they are heavily weighted
toward lower redshifts with a smaller median redshift (0.16) than most
samples.  The median redshift of the RGB Complete sample is essentially the
same as the full RGB sample.

Not surprisingly, many of the lowest redshift objects in the RGB BL\,Lac
sample exhibit a \ion{Ca}{2} break contrast: 13 of the 17 RGB BL\,Lacs with
measurable \ion{Ca}{2} break contrasts lie below the median redshift.  This
implies that earlier samples which required \ion{Ca}{2} break contrasts to be
less than 25\% are missing a significant fraction of low redshift objects, a
result also obtained in the preliminary work on the ``REX'' BL\,Lac sample
\cite[]{rex}.  \cite{slew} noted the typical redshift of a Slew Survey BL\,Lac
is much lower than that of an EMSS object, perhaps indicating the EMSS sample
is incomplete due to misclassification.  Recent observations indeed show that
four EMSS objects should be reclassified as BL\,Lacs based on the
\cite{marcha2} criteria \cite[]{travis}.  These points illustrate the
ambiguity in the current definitions for BL\,Lacs.  The \cite{marcha2}
criteria, adopted in our RGB sample, are however an important step toward a
standard, meaningful definition that is largely independent of orientation and
encompasses the lower luminosity (and presumedly more numerous) objects.

\subsection{The $\alpha_{\rm ro}$ vs.\ $\alpha_{\rm ox}$
Diagram}\label{subsect:aro_aox}
\begin{figure}[t]
\hspace*{0.5in}
\psfig{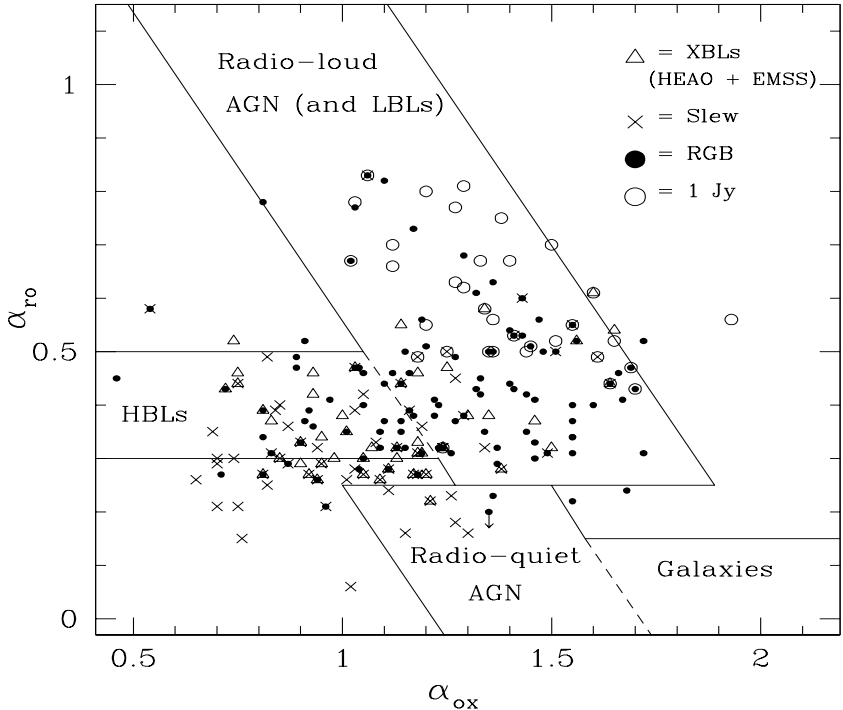}
\caption[]{The $\alpha_{\rm ro}$ vs.\ $\alpha_{\rm ox}$ color-color diagram
for various samples of BL\,Lacs.  The flux densities used to calculate the
spectral indices have all been K-corrected and converted to 5\,GHz (radio
core), 4400\,\AA, and 2\,keV for the three bands.  Objects common to more than
one sample are plotted using the symbols representing all samples to which
they belong.  The EMSS-defined class boundaries are also shown for comparison.
Note that the RGB sample spans both the traditional HBL region (defined by the
EMSS and HEAO samples) and the LBL region (defined by the 1\,Jy
sample).\label{fig:aro_aox}}
\end{figure}

In Figure \ref{fig:aro_aox} we present the $\alpha_{\rm ro}$ vs.\ $\alpha_{\rm
ox}$ color-color diagram for the RGB and comparison BL\,Lac samples.  The
spectral indices for the comparison samples have been recalculated using our
assumptions (\S\ref{rgb_sample}) and based on data in the literature.  When
the redshift is unknown, we use the median redshift for objects in that sample
(see Table \ref{tab:median}).  However, we were not able to correct the
optical fluxes from other samples to include emission from only the AGN
component, as we did for the RGB sample because the \ion{Ca}{2} break
contrasts are generally unavailable.  The correction for host galaxy starlight
is typically an increase (decrease) in $\alpha_{\rm ro}$ ($\alpha_{\rm ox}$)
of 0.06 (0.11) for the 17 ``corrected'' RGB objects which has little effect on
the $\alpha_{\rm ro}$$-$$\alpha_{\rm ox}$ diagram.  Additionally, unlike most
diagrams of this type, the radio flux density used to calculate the
$\alpha_{\rm ro}$ spectral index includes the 5\,GHz flux from only the
arcsecond-scale radio core which best represents the beamed component.  This
is an important point, as extended emission is frequently comparable to core
emission in HBLs \cite[]{lmheao,perlman}.

\placefigure{fig:aro_aox}

Figure \ref{fig:aro_aox} shows the RGB BL\,Lacs exhibit a smooth distribution
in $\alpha_{\rm ox}$, ranging from 0.46$-$1.72, with no hint of the bimodality
which has been previously widely discussed \cite[e.g.,][]{2BLs}.  This result
is in agreement with early results from the ``DXRBS'' \cite[]{dxrbs} and
``REX'' \cite[]{rex} samples which show that a large number of intermediate
objects exists and that there is no clear separation of the HBL and LBL
subclasses (see also \S\ref{subsect:SxSr} below).  This implies the apparent
bimodality could have been caused by selection effects inherent in the two
best-studied, previously known samples: the EMSS and 1\,Jy, a question we 
will examine in greater detail in \S\ref{unified_scheme}.

Both the RGB and Slew survey objects lie in regions of flatter $\alpha_{\rm
ro}$ and steeper $\alpha_{\rm ox}$ than the EMSS or 1\,Jy BL\,Lacs.
Generally, the RGB BL\,Lacs lie along a horizontal band defined by
0.2$<$$\alpha_{\rm ro}$$<$0.6.  All but one of the RGB objects with
$\alpha_{\rm ro}$$>$0.6, also have $\alpha_{\rm ox}$$>$1.0 and therefore have
LBL-like SEDs, although few RGB objects appear to be as extreme as the
majority of 1\,Jy BL\,Lacs.  This likely occurs because any RGB object fainter
than O$\approx$18.5\,mag (the magnitude limit for most of the RGB sample) with
$\alpha_{\rm ro}$$\ge$0.6 must be very radio bright, ${\rm S}_{\rm
r}$$\gtrsim$200\,mJy, a flux density large enough to exclude the majority of
RGB objects.  In contrast, the 1\,Jy sample has an optical magnitude limit of
V$\sim$20\,mag and therefore contains objects with very steep values of
$\alpha_{\rm ro}$.  As the identification of the RGB catalog \cite[]{rgb} is
extended to fainter optical magnitudes, we expect objects with flat
$\alpha_{\rm ox}$ but steep $\alpha_{\rm ro}$ spectral indices (i.e., extreme
LBL-like BL\,Lacs) will be discovered.

\subsection{X-ray to Radio Flux Density Ratios}\label{subsect:SxSr}

Previous BL\,Lac samples show a clear bimodality in the ratio of the X-ray to
radio flux densities of HBLs and LBLs at $\log {\rm S}_{\rm X}/{\rm S}_{\rm
r}$$\simeq$$-$5.5 \cite[]{padovanigiommi,slew,2BLs}.  In contrast, the RGB
BL\,Lacs have a median X-ray to radio logarithmic flux density ratio of
$-5.61$ (Table \ref{tab:median}).  The distribution of the flux ratios (Figure
\ref{fig:SxSr}) shows no evidence for a sharp division between the two
subclasses.  The RGB catalog is therefore the first to contain large numbers
intermediate BL\,Lacs, although hints that these objects existed have been
previously reported by \cite{dxrbs}, \cite{rex} and \cite{rgbIDs}.

Statistical two-sample tests show that the RGB and the RGB Complete samples
are consistent with having been drawn from the same distribution.  This
indicates that the bright optical magnitude limit imposed on the RGB Complete
sample does not drastically affect the overall distribution of S$_{\rm
X}$/S$_{\rm r}$ and, by extension, the fraction of HBLs vs.\ LBLs, a property
for which the different unification models have significantly different
predictions.  Studies of the properties of the Complete Sample can therefore
yield important insights into the origin of the HBL vs.\ LBL subclasses.

\subsection{High Energy Continuum: $\alpha_{\rm xox}$}\label{subsect:axox}
The X-ray emission mechanism in both HBLs and LBLs is most likely dominated by
synchrotron radiation at low energies and inverse-Compton (IC) processes at
higher ($\gamma$-ray) energies \cite[]{BR,GMT,BMU}.  The shape of the X-ray
spectrum is useful for determining at what energy this transition takes place
which is important for understanding the overall energy budget and underlying
jet physics in BL\,Lacs.

While X-ray spectral indices characterize the high energy continuum, the
composite X-ray/optical spectral index, $\alpha_{\rm xox}$$=$$\alpha_{\rm ox}
- \alpha_{\rm x}$ more precisely measures changes in the SED between the
optical and soft X-ray bands by distinguishing the relative importance of IC
and synchrotron emission processes.  If $\alpha_{\rm xox}$$\leq$0, then the
X-rays lie along a powerlaw or steepening synchrotron continuum.  A positive
value of $\alpha_{\rm xox}$ represents a concave spectrum and is likely caused
by a hard IC component.

\begin{figure}[t]
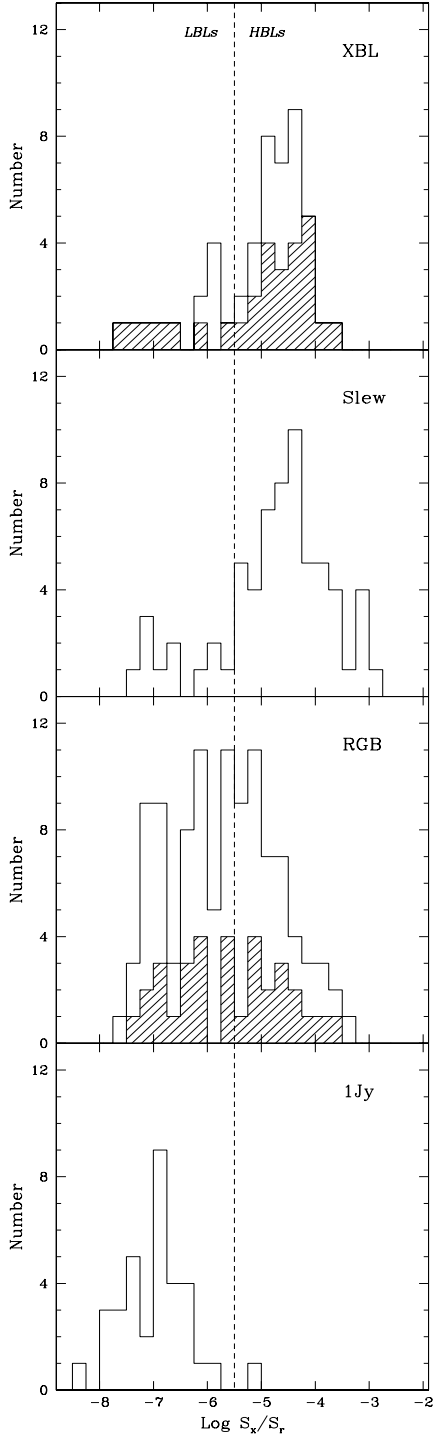

\vspace*{-1.4cm}
\psfig{file=xbl1.SxSr.epsi,height=5.5cm,width=5.9cm}
\vspace*{-0.92cm}
\psfig{file=slew.SxSr.epsi,height=5.5cm,width=5.9cm}
\vspace*{-0.92cm}
\psfig{file=rgb.SxSr.epsi,height=5.5cm,width=5.9cm}
\vspace*{-0.92cm}
\psfig{file=1jy.SxSr.epsi,height=5.5cm,width=5.9cm}
\vspace*{-0.36cm}
\caption[]{The distribution of the logarithm of the X-ray to radio flux
density ratios for various samples of BL\,Lacs {\bf a.} The XBL sample {\bf
b.} The Slew survey sample {\bf c.} The RGB sample {\bf d.} The 1\,Jy sample.
Hatched regions are as described in Figure \ref{zfig}.  Unlike many other
samples, we consider here only the core radio flux density.  The division
between HBLs and LBLs, as defined by the flux ratios of previously known
samples, clearly occurs at  $\log {\rm S}_{\rm X}/{\rm S}_{\rm r}$$=$$-5.5$.
The RGB sample shows no such dichotomy and also includes objects with
traditional HBL- and LBL-like SEDs.\label{fig:SxSr}}
\end{figure}

\placefigure{fig:SxSr}
\clearpage

\begin{figure}[t]
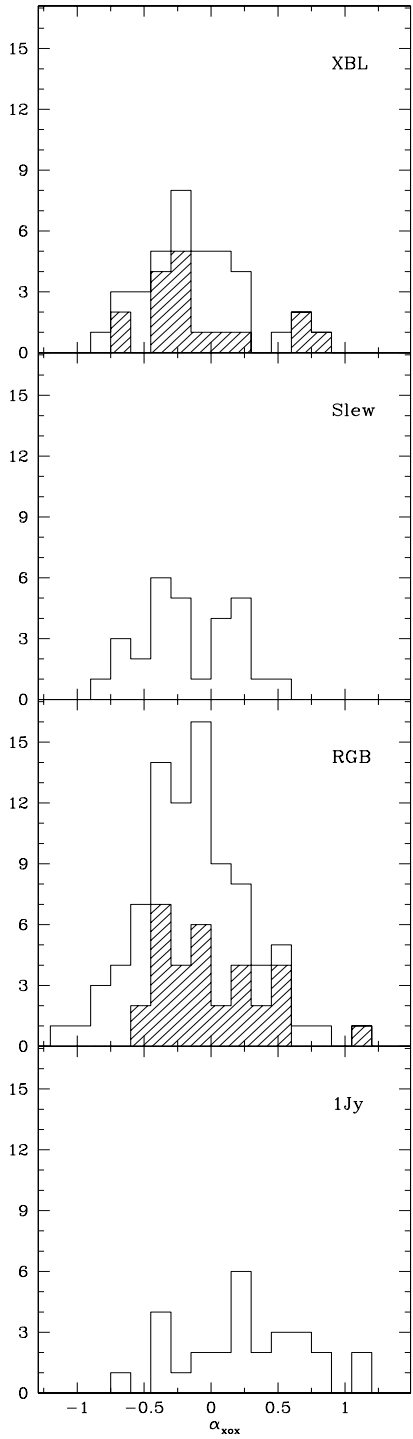

\vspace*{-1.3cm}
\psfig{file=xbl1.axox.epsi,height=5.5cm,width=5.9cm,angle=0}
\vspace*{-0.92cm}
\psfig{file=slew.axox.epsi,height=5.5cm,width=5.9cm,angle=0}
\vspace*{-0.92cm}
\psfig{file=rgb.axox.epsi,height=5.5cm,width=5.9cm,angle=0}
\vspace*{-0.92cm}
\psfig{file=1jy.axox.epsi,height=5.5cm,width=5.9cm,angle=0}
\vspace*{-0.36cm}
\caption[]{The distribution of the difference between the optical--X-ray and
X-ray spectral indices, $\alpha_{\rm xox}$$=$$\alpha_{\rm ox}$$-$$\alpha_{\rm
x}$, for various BL\,Lac samples {\bf a.} The XBL sample {\bf b.} The Slew
survey sample {\bf c.} The RGB sample {\bf d.} The 1\,Jy sample.  Hatched
regions are as described in Figure \ref{zfig}.\label{fig:axox}}
\end{figure}

\placefigure{fig:axox}
\clearpage

If LBLs and HBLs differ solely by orientation, then it should be possible to
explain differences in their SEDs by invoking beaming models and varying only
one free parameter, the angle to the line-of-sight.  However, \cite{rita} find
that there are differences in the SEDs of LBLs and HBLs which are not
attributable to orientation alone: objects with steeper $\alpha_{\rm ro}$
spectral indices (generally LBLs) have a tendency to also have more positive
values of $\alpha_{\rm xox}$.  This indicates the presence of a second,
presumably hard IC, X-ray component and suggests there are intrinsic
differences between LBLs and HBLs which are independent of orientation.

\begin{figure}[t]
\hspace*{0.75in}
\psfig{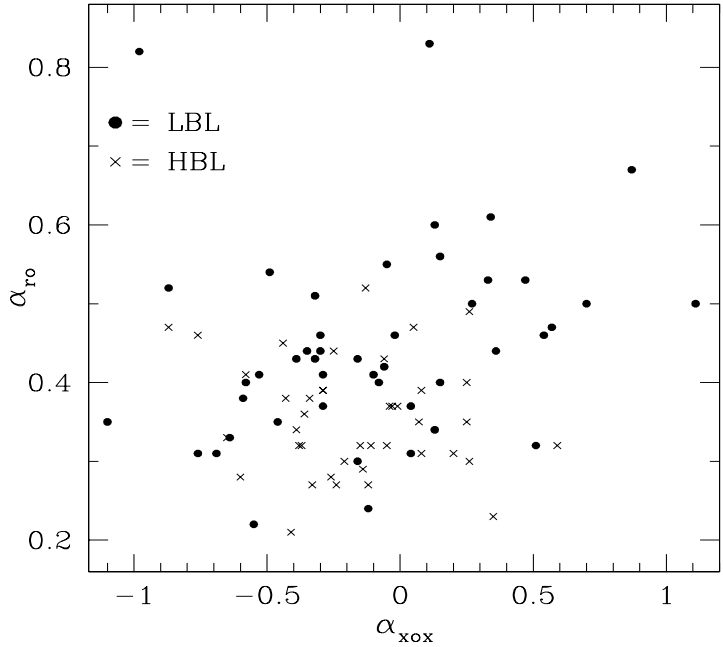}
\caption[]{The $\alpha_{\rm ro}$ vs.\ $\alpha_{\rm xox}$ diagram for the RGB
BL\,Lacs.  Filled circles represent LBLs and X's represent HBLs.  There is a
weak (P$=$92\%) correlation present.  Note that LBLs tend to be the objects
with the steepest $\alpha_{\rm ro}$ and positive (``convex'') $\alpha_{\rm
xox} = \alpha_{\rm ox} - \alpha_{\rm x}$.\label{fig:aro}}
\end{figure}

\placefigure{fig:aro}

Figure \ref{fig:axox} shows the distribution of $\alpha_{\rm xox}$ for the
XBL, Slew, RGB and 1\,Jy samples.  Only those objects with measured X-ray
spectral indices are included (82\% of the RGB sample).  As seen for other
parameters, the RGB sample spans the range exhibited by the extremes of LBLs
and HBLs and many sources exhibit intermediate properties; the RGB Complete
sample exhibits similar characteristics (Table \ref{tab:median}).  To test
the results of \cite{rita}, we searched for a correlation between $\alpha_{\rm
ro}$ and $\alpha_{\rm xox}$ in the RGB sample (Figure \ref{fig:aro}).  Objects
have been divided into HBL- and LBL-like classes based on their X-ray to radio
flux ratios.  There is a correlation with moderate statistical significance
(P$\simeq$92\% using a nonparametric Spearman's $\rho$ statistic) for the RGB
BL\,Lacs.  This is weaker than the correlation found by \cite{rita} who found
P$>$99.99\% using the EMSS and 1\,Jy samples.  However, if we supplement the
intermediate RGB objects with BL\,Lacs from the EMSS, HEAO and 1\,Jy samples,
the probability that a correlation is present increases to $>$99.99\%.  It is
not surprising that the addition of the extreme HBLs (which lie in the bottom
left corner of the diagram) and the extreme LBLs (which lie in the upper right
corner) strengthen the correlation.  This shows that the full range of high
energy spectral shapes exhibited by BL\,Lacs does correlate with BL\,Lac
subclass, confirming the results of \cite{rita}.

\subsection{X-ray $\log$N$-$$\log$S Relationship}\label{subsect:lognlogx}
Both the X-ray and radio $\log$N-$\log$S distributions of the EMSS and 1\,Jy
BL\,Lacs have been studied extensively.  However, as the results presented
here show that BL\,Lacs do not belong to two distinct subclasses, conclusions
drawn from studies of the $\log$N-$\log$S distribution of only the two
extremes of the BL\,Lac population may be misleading.  Nevertheless,
theoretical models based on the 1\,Jy and EMSS data have made interesting
predictions.  Based on the standard orientation-based beamed jet model
unifying FR\,I radio galaxies and BL\,Lacs, the known FR\,I luminosity
function and observational constraints on jet Lorentz factors, \cite{UPS}
calculate the radio radio $\log$N-$\log$S distribution of BL\,Lacs down to
1\,mJy.  We compare the Complete RGB sample to these predictions and other
BL\,Lac samples.

One caveat regarding the completeness of the sample bears mentioning.  We
characterize the RGB BL\,Lacs' radio emission using core flux densities which
can often fall significantly below the total flux density as measured in the
original Green Bank survey.  As a consequence, only those objects with low
radio core-to-lobe ratios populate the lowest radio flux density bins.  (Faint
radio sources with high radio core-to-lobe ratios will have total radio flux
densities below the GB survey flux limit and will be missing from our survey.)
Only the lowest radio flux density bins ($<$20\,mJy based on the GB96 survey
flux limit) are affected by this source of incompleteness.

\begin{figure}[t]
\hspace*{0.75in}
\psfig{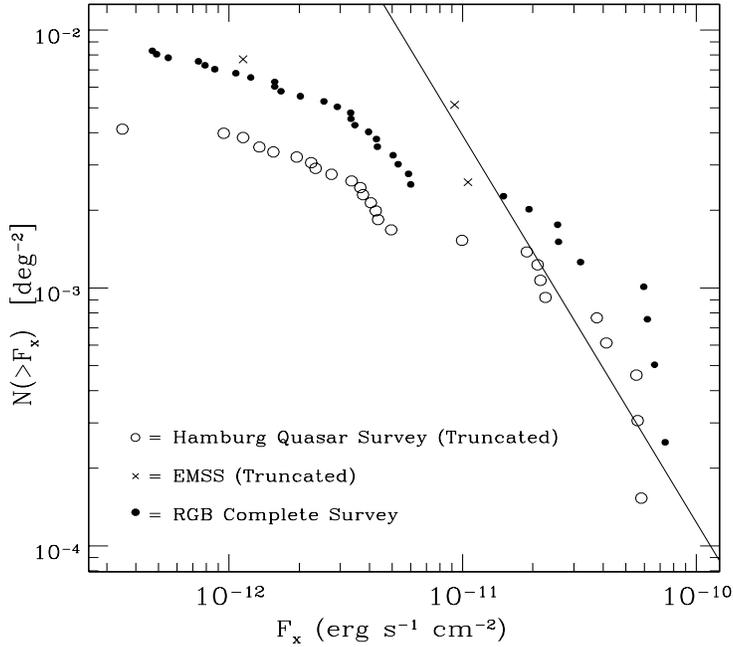}
\caption[]{The X-ray $\log$N-$\log$S relationship for the RGB Complete, EMSS
and HQS samples of BL\,Lacs.  The latter two samples have been truncated to
conform with the optical and radio selection criteria used in the creation of
the RGB Complete sample.  The solid line represents the no-evolution Euclidean
N$\propto$S$^{-3/2}$ powerlaw arbitrarily normalized.\label{fig:lognlogx}}
\end{figure}

\placefigure{fig:lognlogx}

Figure \ref{fig:lognlogx} shows the X-ray $\log$N-$\log$S distribution for the
RGB Complete, EMSS and Hamburg Quasar Survey samples \cite[HQS, ][]{hqs,NA}.
Here we truncate the EMSS and HQS samples so that they match the selection
criteria present in the RGB Complete catalog, namely the objects must be
brighter than O$=$18.0\,mag in the optical and bright enough in the radio to
be present in the GB96 catalog.  This eliminates 19 of the 22 EMSS objects,
while the HQS sample is diminished from 61 objects to 27.

Also shown in the figure is the Euclidean no-evolution model defined by
N($>$F$_{\rm X}$)$\propto$F$^{-3/2}_{\rm X}$ and arbitrarily normalized to
N($>$F$_{\rm X}$)=2.0$\times 10^{-3}$ at F$_{\rm
X}$$=$1.5$\times$10$^{-11}$\,erg\,s$^{-1}$cm$^{-2}$.  All three samples fall
well below the Euclidean relationship at faint flux levels.  The RGB catalog
exhibits a higher density of objects than the HQS given identical flux limits.
At a flux of $\sim$10$^{-12}$\,erg\,s$^{-1}$cm$^{-2}$, the HQS number density
is $\sim$1.7 times lower than the RGB sample likely a result of the very
restrictive $\alpha_{\rm ox}$$<$1.1 ($\log$(S$_{\rm X}$/S$_{\rm o}$)$>$1.3)
HQS selection criterion (see Figure \ref{fig:aro_aox}).

The X-ray $\log$N-$\log$S distribution shows slight evidence for the ``bump''
at $\sim$$3\times 10^{-11}$\,erg\,s$^{-1}$cm$^{-2}$ reported both by
\cite{Xbump} and \cite{NA}, although the number statistics are poor.  There
also appears to be a rise in the surface density near F$_{\rm
X}$$=$$8\times10^{-12}$\,erg\,s$^{-1}$cm$^{-2}$.  The number statistics are
better here, but it is still difficult to assess the significance of this
feature.  The distribution could alternately be characterized by a powerlaw
with a deficit of sources near $5-10\times10^{-12}$\,erg\,s$^{-1}$cm$^{-2}$.

\cite{badeComplete} have noted that the X-ray $\log$N-$\log$S distribution of
HQS BL\,Lacs is more monotonic when extreme HBLs ($\alpha_{\rm ox}$$<$0.91)
are excluded.  From this, \cite{badeComplete} infer that intermediate BL\,Lacs
constitute the ``basic BL\,Lac population'' and that X-ray dominated BL\,Lacs
are those objects which are observed in a state of enhanced X-ray activity.
In order to test this hypothesis, we examined the RGB X-ray $\log$N-$\log$S
distribution, excluding first those RGB BL\,Lacs with $\alpha_{\rm
ox}$$<$0.91 as was done in \cite{badeComplete}.  This criterion eliminates
only two objects from the Complete sample, so the overall shape is not
affected and the feature at F$_{\rm
X}$$=$$8\times10^{-12}$\,erg\,s$^{-1}$cm$^{-2}$ remains.  Even if we raise
the cutoff to values of $\alpha_{\rm ox}$ to 1.0, 1.1 or 1.2, the feature at
F$_{\rm X}$$\approx$$8\times10^{-12}$\,erg\,s$^{-1}$cm$^{-2}$ remains although
the number statistics necessarily deteriorate.  This contradicts the result of
\cite{badeComplete}, although we cannot rule it out because of the unknown
effects of the optical magnitude limit on the RGB Complete sample.
Therefore we find the feature at F$_{\rm
X}$$\approx$$8\times10^{-12}$\,erg\,s$^{-1}$cm$^{-2}$ is significant, and 
does not seem to depend on properties of only an extreme subset of BL\,Lacs.
We believe it therefore does not indicate extreme HBLs are objects in a state
of enhanced X-ray activity, but suggest it is an indication of the type of
breaks which are expected in $\log$N-$\log$S distributions of beamed objects
(see below).

\subsection{Radio $\log$N$-$$\log$S Relationship}\label{subsect:lognlogs}
\begin{figure}[t]
\hspace*{0.75in}
\psfig{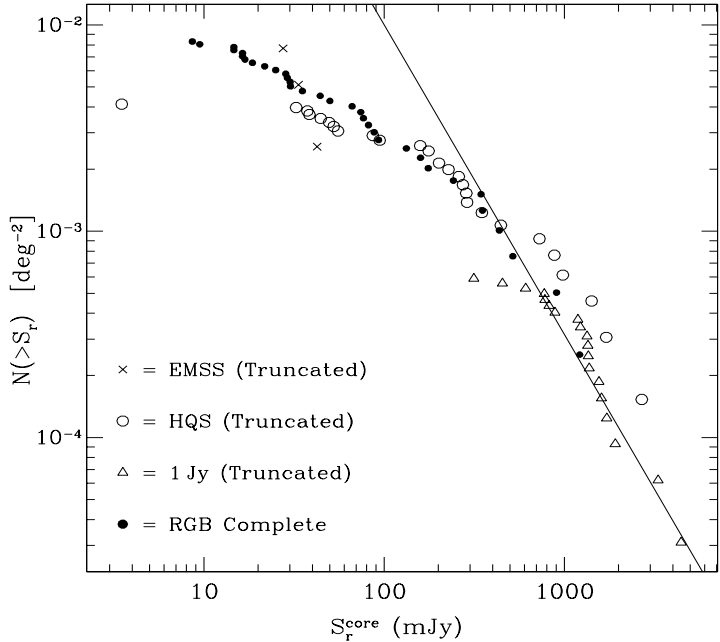}
\caption[]{The radio $\log$N-$\log$S relationship for the RGB Complete, EMSS,
HQS and 1\,Jy samples of BL\,Lacs.  The latter three samples have been
truncated to conform with the selection criteria used in the creation of the
RGB Complete sample. The solid line represents the no-evolution Euclidean
N($>$S)$\propto$S$^{-3/2}$ powerlaw arbitrarily
normalized.\label{fig:lognlogr.trunc}}
\end{figure}

Figure \ref{fig:lognlogr.trunc} shows a radio $\log$N-$\log$S distribution
with the RGB, EMSS, HQS and 1\,Jy samples included.  As before, all samples
have been truncated to include only those objects which adhere to the RGB
selection criteria.  This has the effect of reducing the 34 object 1\,Jy
complete sample to 19 objects.  Note also that for all samples we plot core
rather than total radio flux densities because the total emission contains 
kiloparsec-scale emission which is not as intimately connected with the
relativistic flow in the center of the AGN.  Many studies do not make this
core vs.\ total flux distinction, but considering the wide range of
core-to-lobe ratios exhibited by BL\,Lacs, we feel it important to do so here.

\placefigure{fig:lognlogr.trunc}

Several points are evident from the figure.  The incompleteness of the 1\,Jy
sample below radio core flux densities of $\sim$1\,Jy is apparent from the
turnover in the number counts in the diagram.  However, above $\sim$700\,mJy
the HQS, RGB and (to within $\sim$20\%) the 1\,Jy samples agree well,
indicating that the effects of the X-ray flux and optical magnitude limits do
not do not severely affect the RGB Complete sample at these high radio flux
density limits.  In the middle part of the diagram ($\sim$70$-$700\,mJy), the
RGB and HQS samples are roughly consistent with each other, unlike the results
obtained for the X-ray $\log$N-$\log$S distribution.  At the faintest radio
flux densities, the surface density of HQS BL\,Lacs falls $\sim$2 times below
the RGB sample, indicating selection criteria based on identifying optically
bright sources with flat values of $\alpha_{\rm ox}$ produce an incompleteness
that increases with decreasing radio flux density.

\begin{figure}[t]
\hspace*{0.75in}
\psfig{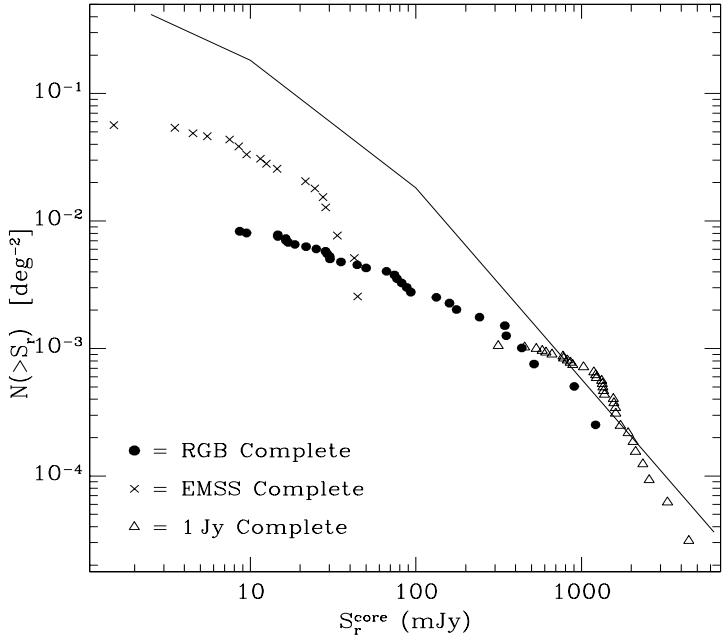}
\caption[]{The radio $\log$N-$\log$S relationship for the RGB Complete, EMSS
and 1\,Jy samples of BL\,Lacs.  The solid line represents the theoretical
beaming model prediction of Urry, Padovani \& Stickel (1991;
\S5.6).\label{fig:lognlogr}}
\end{figure}

\placefigure{fig:lognlogr}

While it may be tempting to infer the evolutionary behavior of BL\,Lacs from
the curvature of the RGB $\log$N-$\log$S distributions, it is dangerous to do
so without careful modeling of the effects introduced by the X-ray and optical
magnitude flux limits.  This is illustrated in Figure \ref{fig:lognlogr} which
shows the radio $\log$N-$\log$S distribution for the RGB Complete sample in
comparison with the full (not truncated) 1\,Jy and complete EMSS samples.
Here the effects of the radio, optical and X-ray limiting flux densities
inherent in the RGB catalog are clearly evident, particularly at low radio
flux densities where the counts flatten appreciably, falling a factor of
$\sim$4 below the EMSS sample, which itself represents only a fraction of the
radio faint BL\,Lac population.

Figure \ref{fig:lognlogr} also shows a theoretical beaming model for the
$\log$N-$\log$S distribution (see and Urry and Shafer 1984 and Urry et al.\
1991 for details).  \nocite{US} \nocite{UPS}  This model is based on a
simulation of a randomly oriented parent population of FR\,I radio galaxies
which follow an assumed cosmological evolution model and whose jets have a
powerlaw distribution of bulk jet Lorentz factors.  The resulting BL\,Lac
luminosity function is characterized by a double powerlaw, flattened by
beaming at low radio powers with a break that is shifted to higher radio
powers relative to breaks in the luminosity function of the parent population.
When converted to surface number densities and fluxes, the $\log$N-$\log$S
distribution of BL\,Lacs exhibits a Euclidean slope ($-$3/2) down to 100\,mJy,
flattens to N($>$S$_{\rm r}$)$\propto$S$_{\rm r}^{-1}$ between 10$<$S$_{\rm
r}$$<$100\,mJy, and then becomes very flat ($\propto$S$_{\rm r}^{-0.6}$) for
S$_{\rm r}$$<$10\,mJy.  Beaming therefore tends to flatten source counts to an
increasing degree at lower flux densities.

Unfortunately, because neither the EMSS nor RGB samples are truly complete
(containing {\em all\/} BL\,Lacs) at radio flux densities below $\sim$1\,Jy,
it is not possible to directly compare the specific predictions of this model
with any of the samples in Figure \ref{fig:lognlogr} other than the 1\,Jy
which provides the normalization.  However, we can make some general
observations.  The RGB sample follows a Euclidean slope above $\sim$200\,mJy.
At intermediate radio flux densities, it flattens more than the model but, as
stated above, this is at least partly a result of the incompleteness caused by
the RGB Complete sample's flux limits.  At $\sim$40\,mJy, the surface density
of the RGB and EMSS samples coincide, but below $\sim$20\,mJy, the Complete
EMSS (HBL) sample quickly overwhelms the RGB sample.  Nevertheless, the EMSS
sample falls well below the prediction of \cite{UPS}.  It is this deficit of
low flux/luminosity BL\,Lacs which led \cite{morris} to postulate that HBLs
may evolve negatively, a claim reinforced by \cite{travis} in their new
analysis of the V/V$_{\rm max}$ analysis of an updated Complete EMSS sample.
However, the EMSS survey consists of only the most extreme X-ray$-$dominated
objects.  As discussed below, we believe the dichotomy of HBLs and LBLs is
a result of selection effects and conclusions drawn about the evolution of
only one extreme end of the distribution should therefore be treated
cautiously.

\section{Discussion}\label{unified_scheme}
\subsection{Comparison with Previous BL\,Lac Surveys}

The principal result of this study is that the RGB BL\,Lacs exhibit a 
continuous range in SEDs rather than segregating into two distinct classes:
HBLs vs.\ LBLs (or XBLs vs.\ RBLs).  Clearly the RGB sample smoothly spans a
vast range of SEDs: 5 orders of magnitude in S$_{\rm x}$/S$_{\rm r}$,
0.5$<$$\alpha_{\rm ox}$$<$1.7 and 0.2$<$$\alpha_{\rm ro}$$<$0.8.  Most RGB
BL\,Lacs have intermediate properties (e.g., S$_{\rm x}$/S$_{\rm r}$ ratios)
between the HBL and LBL extremes that dominated earlier samples.  Similar
preliminary results have also been reported for the deep DXRBS and REX X-ray
surveys \cite[]{dxrbs,rex}.  The question remains whether these results are an
accurate representation of the BL\,Lac population as a whole, or whether the
true BL\,Lac distribution is bimodal and the flux limits of the RGB survey
unfortuitously tuned to make make it particularly sensitive to objects with
intermediate SEDs.  Occam's Razor makes it tempting to hypothesize that the
intermediate nature of the RGB sample is an accurate reflection of an
underlying unimodal distribution and that the previously observed bimodal
distribution is a result of high flux limits of previous surveys, but this
hypothesis bears closer examination.

\begin{figure}[t]
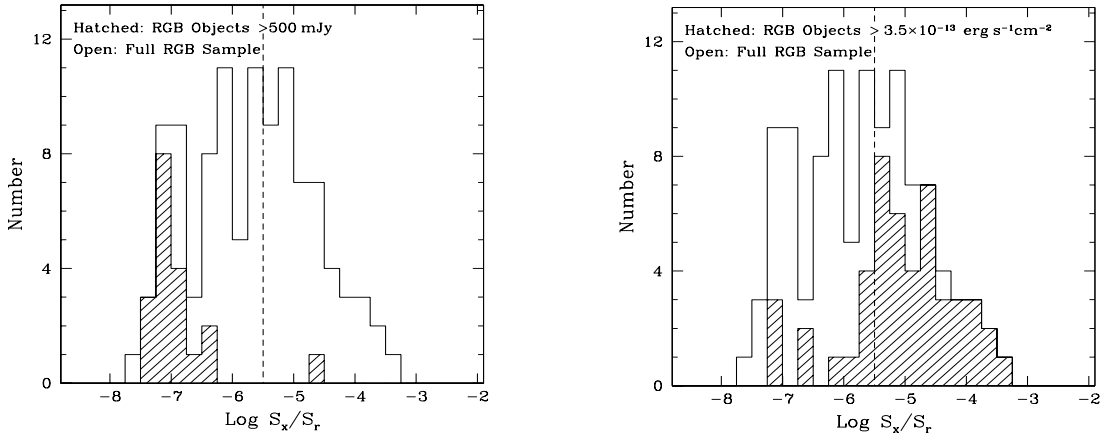

\psfig{file=trunc500.epsi,height=6.0cm,width=6.7cm}
\vspace*{-6.0cm}
\hspace*{8.0cm}
\psfig{file=trunc.3_5.epsi,height=6.0cm,width=6.7cm}
\caption[]{Comparison of the $\log {\rm S}_{\rm X}/{\rm S}_{\rm r}$ ratio for
the full RGB sample (upper solid line) and those objects with {\bf a.} radio
flux densities greater than 500\,mJy (hatched histogram) or {\bf b.} X-ray
fluxes greater than $3.5\times10^{-12}$\,erg\,s$^{-1}$cm$^{-2}$.  Note the
similarity between the hatched distributions and those in Figure
4.\label{fig:truncated}}
\end{figure}

\placefigure{fig:truncated}

Figure \ref{fig:truncated} shows what the RGB's distribution of S$_{\rm
x}$/S$_{\rm r}$ ratios would have been had either the radio (Fig.\
\ref{fig:truncated}a) or X-ray (Fig.\ \ref{fig:truncated}b) flux limits been
higher.  Not surprisingly, increasing the radio flux limit preferentially
selects increasingly more radio-dominant BL\,Lacs.  At a limiting radio flux
density of 500\,mJy a distribution roughly consistent with that of the 1\,Jy
sample (Fig.\ \ref{fig:SxSr}) results.  (Too few RGB objects are brighter 
than 1\,Jy to make a meaningful comparison at this higher radio flux density
limit.)  Direct comparisons with X-ray$-$selected samples are complicated
since none of the comparison surveys were conducted over the same energy band
as the RGB (0.1$-$2.4\,keV with the ROSAT PSPC) and X-ray flux limits are
sensitive to the assumed photon index and Galactic column density.  In
addition, the Slew Survey has a factor of several range of limiting
sensitivities while the HEAO-1 sample is a result of a flux density (not flux)
limited survey complicating a precise comparison with the RGB sample.

Despite these ambiguities, increasing the limiting X-ray flux of the RGB
survey selects the more X-ray$-$dominated RGB BL\,Lacs and at a flux limit of
$3.5\times10^{-12}$\,erg\,s$^{-1}$cm$^{-2}$, the distribution is roughly
consistent with both the HEAO-1 and Slew Survey distributions (Figs.\
\ref{fig:SxSr}a and \ref{fig:SxSr}b).  However, the effective limit of the
EMSS in the ROSAT band is $2.2 - 3.5 \times 10^{-2}$\,$\mu$Jy (T.\ Rector,
private communication) which is only marginally brighter than the RGB X-ray
flux limit.  Therefore if our assumption of a unimodal S$_{\rm x}$/S$_{\rm r}$
ratio for BL\,Lacs is correct, then we predict (based on only a small
difference in the RGB and EMSS X-ray survey flux limits), that the EMSS sample
should closely resemble that of the RGB; this clearly is not the case.
However, \cite{travis} have made a careful re-examination of the EMSS
identifications and have found several new BL\,Lacs that were misidentified
because of their strong 4000\,\AA\ break contrasts.  The addition of these
sources does bring the EMSS and RGB S$_{\rm x}$/S$_{\rm r}$ distributions into
somewhat closer agreement.  Also, a new analysis by \cite{d40} which extends
the complete EMSS sample down to $2.0\times 10^{-13}$\,erg\,s$^{-1}$cm$^{-2}$
and declinations $>$40$^{\circ}$, also preferentially adds objects with
intermediate S$_{\rm x}$/S$_{\rm r}$ ratios, but not sufficiently to make the
RGB and EMSS distributions consistent.  Nevertheless, agreement between the
1\,Jy, HEAO-1 and Slew samples with that of the RGB sample were either its
radio or, respectively, X-ray flux limits raised, is compelling.

\begin{figure}[t]
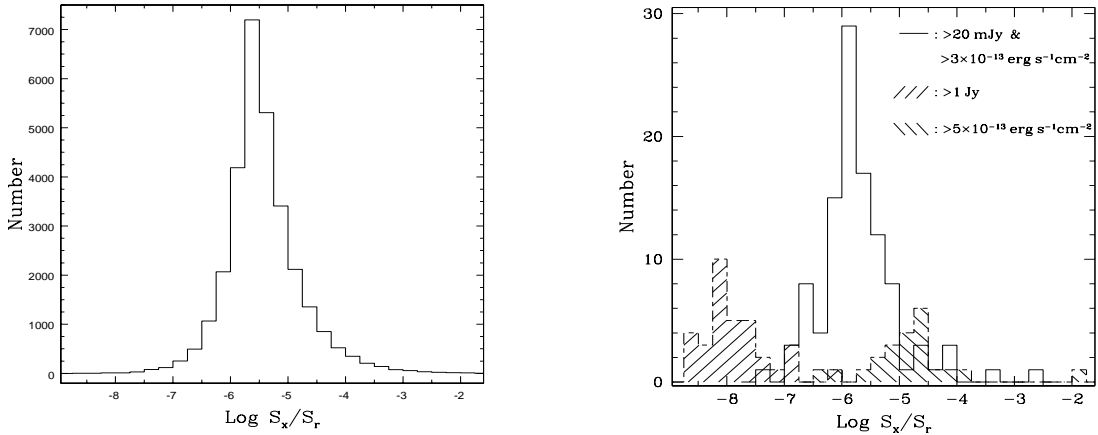

\psfig{file=monte.epsi,height=6.0cm,width=6.7cm}
\vspace*{-6.0cm}
\hspace*{8.0cm}
\psfig{file=monte.samp.epsi,height=6.0cm,width=6.7cm}
\caption[]{Results of a Monte Carlo simulation of the $\log {\rm S}_{\rm
X}/{\rm S}_{\rm r}$ for BL\,Lacs.  The model shown assumes the $\log
N$$-$$\log S$ relationship follows a S$^{-1.2}$ powerlaw in the radio and a
S$^{-0.8}$ powerlaw in the X-ray.  {\bf a.} The underlying distribution which
is unimodal and peaks at $\log {\rm S}_{\rm X}/{\rm S}_{\rm r}$$=$$-5.6$ {\bf
b.} The distributions which would result from this sample if surveys with flux
limits of 1\,Jy (right-slanted hatched histogram), $5 \times
10^{-13}$\,erg\,s$^{-1}$cm$^{-2}$ (left-slanted hatched histogram) or dual
flux limits of 20\,mJy and $3 \times 10^{-13}$\,erg\,s$^{-1}$cm$^{-2}$ were
performed.  Flux limits have been chosen based on the limits inherent in the
1\,Jy, EMSS and RGB surveys, respectively and the total number of objects in
these simulated samples have been matched to the number of objects in the
survey they are intended to represent.  Although the agreement is not perfect,
the similarity with Figure 4 is suggestive that the $\log {\rm S}_{\rm X}/{\rm
S}_{\rm r}$ distribution for BL\,Lacs is unimodal and the bimodal distribution
reflected in the EMSS and 1\,Jy samples is a result of the flux limits of
these different surveys.\label{fig:monte}}
\end{figure}

Another demonstration that the RGB survey likely accurately reflects an
underlying unimodal distribution arises from the following Monte Carlo
simulation.  We create hypothetical populations of BL\,Lacs where both radio
and X-ray $\log N$$-$$\log S$ distributions obey simple powerlaws, and radio
and X-ray fluxes are independent variables.  The first assumption is clearly
an oversimplification since there is ample evidence for breaks in both the
radio and X-ray $\log N$$-$$\log S$ relationships (including this paper) but
including these breaks would introduce many new parameters to the model
(location of the break(s) and the slopes above an below them) without adding
substantially to the general sense of the results.  Figure \ref{fig:monte}
shows a simulation of 30,000 sources whose radio and X-ray fluxes are
constrained to be $>$1\,mJy and $>$$5 \times
10^{-14}$\,erg\,s$^{-1}$cm$^{-2}$, respectively.  These limits were chosen
because they are fainter than either the radio or X-ray flux of any BL\,Lac in
any of the relevant comparison samples (EMSS, RGB or 1\,Jy).  We then applied
the different flux limits of the comparison surveys and randomly chose objects
in order to create samples with the same number of objects.  The particular
simulation shown assumed the $\log N$$-$$\log S$ powerlaw slopes were $-1.2$
and $-0.8$ in the radio and X-ray, respectively.  (See Urry, Padovani \&
Stickel 1991 and Maccacaro et al.\ 1984 for the relevance of these specific
indices.)\nocite{UPS,lognlogx}  

\placefigure{fig:monte}

Clearly Figure \ref{fig:monte} suggests the underlying distribution is
unimodal; this is, in fact, a universal among all the simulations, although
the location of the peak in the distributions varies with the radio and X-ray
logN$-$logS slopes.  The distribution shown in Figure \ref{fig:monte} peaks at
$\log$S$_{\rm x}$/S$_{\rm r}$$=$$-5.61$.  The distributions resulting from
flux limits at 1\,Jy ($5 \times 10^{-13}$\,erg\,s$^{-1}$cm$^{-2}$) are
acceptably consistent with the distributions shown in Figure \ref{fig:SxSr}
for the 1\,Jy (EMSS) samples while the sample created by enforcing a radio
flux limit of 20\,mJy and an X-ray flux limit of $3 \times
10^{-13}$\,erg\,s$^{-1}$cm$^{-2}$ is an acceptable match to the RGB sample's
distribution, given the simplicity of our assumptions.

Both the above analyses strongly point towards the simple conclusion that the
true distribution of BL\,Lac SEDs, and S$_{\rm x}$/S$_{\rm r}$ ratios in
particular, is unimodal and accurately represented by the RGB survey.   Our
results also show that the previously observed bimodality can be explained as
a result of naturally occurring observational selection effects present in
older surveys.  A similar conclusion is reached by \cite{rex} in their
preliminary examination of the REX survey BL\,Lacs and in the theoretical
modeling of \cite{bolometricL}.

\subsection{A Unified BL\,Lac Population}
Differences between the various subclasses of BL\,Lacs have been attributed
to different jet orientations to the line-of-sight \cite[e.g.,
][]{vista,unified}.  However, further analysis showed that the range of SEDs
observed in LBLs and HBLs cannot be reproduced by simple changes in jet
orientation \cite[e.g.,][]{rita,bolometricL} and our results confirm this
(\S\ref{properties}).  \cite{padovanigiommi} propose an alternative to this
orientation-based model, namely, that LBLs and HBLs essentially share the same
range in orientation, but have intrinsically different SEDs.  They postulate
that the frequency at which the synchrotron break occurs, $\nu_{\rm break}$,
differs intrinsically in LBLs and HBLs.  This then alters the balance of
synchrotron and inverse Compton emission present in soft X-rays, moving the
objects in the $\alpha_{\rm ro}$$-$$\alpha_{\rm ox}$ diagram.  This model
predicts that BL\,Lacs whose spectra break at high frequencies should be
HBL-like and lie along the horizontal path labeled ``HBLs'' in Figure
\ref{fig:aro_aox}.  Objects with $\nu_{\rm break}$ at low frequencies should
lie along the diagonal swath labeled ``LBLs'' in the $\alpha_{\rm
ro}$$-$$\alpha_{\rm ox}$ plane.  Assuming $\langle\alpha_{\rm
ox}\rangle$$\simeq$0.7 and $\langle\alpha_{\rm ro}\rangle$$\simeq$0.37, as
derived from the EMSS sample, and a spectral index $\alpha_{\rm
break}$$\simeq$1.9, valid for $\nu$$>$$\nu_{\rm break}$, \cite{padovanigiommi}
were able to reproduce the bimodal locations of the EMSS XBLs and 1\,Jy RBLs
in the color-color diagram.  If this model is correct, then all new BL\,Lacs
should continue to reside along constrained loci in the $\alpha_{\rm
ro}$$-$$\alpha_{\rm ox}$ plane, although this SED-based model need not lead to
a bimodal separations of classes.  Our results show that this may be the case
with the RGB objects following the traditional paths in the $\alpha_{\rm ro}$
vs.\ $\alpha_{\rm ox}$ plane, although the scatter is significant and the
triple-flux limits restrict the range of SEDs to which we are sensitive.

Two additional new unification models attempt to explain the physical origin
of the wide range of SEDs in BL\,Lacs.  The first asserts that the shape of
the SED is linked to the bolometric luminosity \cite[]{bolometricL}.  In this
model, HBLs and LBLs are manifestations of the same phenomenon, but the
broadband SED varies in a predictable way with bolometric luminosity.  This
model can reproduce the disparate distribution of the 1\,Jy and Slew Survey
samples in the $\alpha_{\rm ro}$$-$$\alpha_{\rm ox}$ plane, and predicts an
intrinsically smooth transition between the two extremes.  The other model
predicts that the origin of the SED differences lies in differences in the
electron kinetic luminosity of the jet which is related to jet size
\cite[L$_{jet} \propto r^2$;][]{kineticjet}.  This model also predicts a
smooth transition between the previously disparate BL\,Lac subclasses which
qualitatively agrees with the distribution of RGB objects.  Quantitative
comparison of these models with our data is, however, difficult both because
the RGB sample is triply flux limited and because the RGB Complete sample's
limiting magnitude of 18.0 curtails the range of possible SEDs to which the
sample is sensitive.  We are however currently attempting to model the
selection effects, but extending the Complete sample to a fainter magnitude
limit would clearly alleviate some restrictions.

The wide range of spectral indices exhibited by the RGB BL\,Lacs has important
implications for followup surveys for RASS BL\,Lacs.  Figure \ref{fig:aro_aox}
shows not only the distribution of BL\,Lacs in the $\alpha_{\rm
ro}$$-$$\alpha_{\rm ox}$ diagram, but also the color-color classification
boundaries established in the EMSS by \cite{stocke}.  These regions have been
used to successfully identify new BL\,Lac objects.  Nearly all Einstein Slew
Survey objects in the region defined by $\alpha_{\rm ro}$$=$(0.3,0.6) and
$\alpha_{\rm ox}$$=$(0.55,1.2) were spectroscopically confirmed as BL\,Lacs
\cite[]{schachter,slew}.  A criterion of $\alpha_{\rm ox}$$<$1.1 was used to
select candidates for the Hamburg Quasar Survey BL\,Lac sample
\cite[]{NA}.  However, the RGB sample, selected without any spectral index
criteria, shows that targeting only those objects in the RASS with particular
color-color indices will miss a large fraction of the BL\,Lacs.  If, for
example, the EMSS BL\,Lac class boundaries had been used to select candidates,
67\% of the RGB BL\,Lacs would have been excluded (70\% of the sources in our
Complete sample).  Using the criterion in \cite{NA} would have excluded 70\%
of our objects (and 76\% of objects in the Complete sample).  Therefore while
the likelihood of finding BL\,Lacs does increase dramatically with decreasing
$\alpha_{\rm ox}$, candidates cannot be chosen based on their SEDs without
producing highly biased samples leading to incorrect or uncertain conclusions.

Finally, because we lack detailed knowledge of the effects of the triple flux
limits on the characteristics of the entire RGB BL\,Lac sample, we are not at
present able to definitively evaluate the predictions of any of the
FR~I/BL\,Lac or HBL/LBL unification models.  Some insights might be obtained
with Monte Carlo simulations which exclude sources which fall below our RGB
flux limits, but this, of course, requires a more elaborate simulation of
the intrinsic properties of the BL\,Lac population than the simple simulation
shown in Figure \ref{fig:monte}.  Two additional approaches can
be pursued.  First, redshifts for the RGB Complete Sample can be obtained with
currently available optical telescopes.  This would transform the source
counts into correctly normalized luminosity functions, which could then be
compared with luminosity functions of FR\,I radio galaxies.  But here again,
the selection effects must be precisely taken into account.  Second, the RGB
sample can be extended to reduce the effects of the flux limits.  Most
importantly, spectra should be obtained for RGB objects fainter than
$\sim$18.5 in order to discriminate BL\,Lacs from other X-ray/radio sources.
While improvements in the X-ray flux limit the RGB catalog are not likely in
the near future, the NVSS \cite[]{nvss} and FIRST \cite[]{first} radio surveys
permit a factor of $\simeq$20 reduction in the radio flux limit.  BL\,Lac
samples based on these radio surveys are currently being pursued
\cite[]{rex,firstbllacs}.  Here again, substantial spectroscopy of optically
faint objects will be needed.  Significant progress in refining BL\,Lac
unification scenarios thus depends on spectroscopy with new 8-meter class
telescopes, which fortunately are proliferating today.

\acknowledgments
We wish to thank Nahum Arav, Michael Brotherton and the referee, Simon Morris,
for a variety of useful comments which greatly improved the final
manuscript.  This work was partially supported by NASA under Grant NAGW-2120
to EDF and partially by the Department of Energy at the Lawrence Livermore
National Laboratory under contract W-7405-ENG-48.  RIK acknowledges support by
Fermi National Accelerator Laboratory under the U. S. Department of Energy
contract No.\ DE-ACO2-76CH03000.  We have made use of the NASA/IPAC
Extragalactic Database, operated by the Jet Propulsion Laboratory, California
Institute of Technology, under contract with NASA.  SALM also acknowledges
partial support from the NASA Space Grant Consortium through their Space Grant
Fellow program and the NSF (grant AST-98-02791).

\newpage


\bibliography{bllacs}
\bibliographystyle{natbib}

\clearpage

\newpage

\begin{figure}
\vspace*{1.55in}
\hspace*{-0.75in}
\psfig{file=bllac.allinfo.002,angle=90}
\end{figure}

\begin{figure}
\vspace*{1.55in}
\hspace*{-0.75in}
\psfig{file=bllac.allinfo.003,angle=90}
\end{figure}

\begin{figure}
\vspace*{1.55in}
\hspace*{-0.75in}
\psfig{file=bllac.allinfo.004,angle=90}
\end{figure}

\begin{figure}
\vspace*{1.55in}
\hspace*{-0.75in}
\psfig{file=bllac.allinfo.005,angle=90}
\end{figure}

\begin{figure}
\vspace*{1.55in}
\hspace*{-0.75in}
\psfig{file=bllac.allinfo.006,angle=90}
\end{figure}

\clearpage

\begin{figure}
\vspace*{1.25in}
\hspace*{-0.75in}
\psfig{file=bllac.allinfo.007,angle=90}
\end{figure}

\begin{figure}
\vspace*{1.25in}
\hspace*{-0.75in}
\psfig{file=bllac.allinfo.008,angle=90}
\end{figure}

\begin{figure}
\vspace*{1.25in}
\hspace*{-0.75in}
\psfig{file=bllac.allinfo.009,angle=90}
\end{figure}

%


\end{document}